\documentstyle[pre,aps,epsf]{revtex}
\begin{document}
\def\be{\begin{equation}}
\def\ee{\end{equation}}
\def\bc{\begin{center}}
\def\ec{\end{center}}
\def\bea{\begin{eqnarray}}
\def\eea{\end{eqnarray}}

\title{Off equilibrium response function in the one dimensional 
random field Ising model}
\author{F.~Corberi$^{1,\dag}$, A.~de~Candia$^{2,\diamond}$, 
E.~Lippiello$^{1,\ddag}$ and M.~Zannetti$^{1,\S}$}
\address {$^1$ Istituto Nazionale di Fisica della Materia, Unit\`a
di Salerno and Dipartimento di Fisica ``E.Caianiello'', 
Universit\`a di Salerno,
84081 Baronissi (Salerno), Italy\\
$^2$ Istituto Nazionale di Fisica della Materia, Unit\`a
di Napoli and Dipartimento di Scienze Fisiche, Universit\`a di Napoli
``Federico II'', via Cintia, 80126 Napoli, Italy}
\maketitle
\begin{abstract}

A thorough numerical investigation of the slow dynamics in the $d=1$
random field Ising model in the limit of an infinite ferromagnetic
coupling is presented. Crossovers from the preasymptotic pure regime
to the asymptotic Sinai regime are investigated for the average
domain size, the autocorrelation function and staggered magnetization.
By switching on an additional small random field at the time $t_w$
the linear off equilibrium response function is obtained, which 
displays as well the crossover from the nontrivial behavior of the $d=1$ pure 
Ising model to the asymptotic behavior where it vanishes identically.

\end{abstract}
\bigskip

\dag corberi@na.infn.it $\diamond$ decandia@na.infn.it

\ddag lippiello@sa.infn.it \S zannetti@na.infn.it


\section{Introduction}

A phase ordering system, such as a ferromagnet quenched below the
critical point, offers the simplest example of  slow relaxation 
with many of the interesting features observed also in more complex 
glassy systems\cite{Bouchaud97}. At
the core of this phenomenology is the separation of the time scales
of fast and slow variables. In a domain forming system equilibrium
is rapidly reached in the interior of domains while the interfacial
degrees of freedom remain out of equilibrium for a time which diverges
with the size of the system. In these conditions the order parameter
autocorrelation function splits into the sum
\be
G(t,t_w) = G_{\text{st}}(t-t_w) + G_{\text{ag}}(t/t_w)
\label{i1}
\ee
where $G_{\text{st}}(t-t_w)$ is the stationary time translation invariant (TTI)
contribution due to fluctuations in the bulk of domains and 
$G_{\text{ag}}(t/t_w)$ is
the aging, or scaling contribution\cite{Furukawa89,Bray94} originating in 
the off equilibrium fluctuations. In the time
scale over which $G_{\text{st}}(t-t_w)$ decays to zero 
$G_{\text{ag}}(t/t_w)$ stays practically constant at $M^2$, where $M$
is the equilibrium value of the order parameter, while the decay of
$G_{\text{ag}}(t/t_w)$  takes place for much larger time separations.
Using $G(t,t)=1$ for spin variables, this implies $G_{\text{st}}(0)=1-M^2$.

A structure similar to~(\ref{i1}) is observed also in the linear 
response at the time $t$ to
a random external field switched on at the earlier time $t_w$
\be
\chi (t,t_w) = \chi_{\text{st}}(t-t_w) + \chi_{\text{ag}}(t,t_w) 
\label{i2}
\ee
where the stationary contributions $\chi_{\text{st}}(t-t_w)$ 
and $G_{\text{st}}(t-t_w)$ are related by the
equilibrium fluctuation dissipation theorem (FDT)
\be 
T\chi_{\text{st}}(t-t_w)=G_{\text{st}}(0)-G_{\text{st}}(t-t_w)
\label{FDT}
\ee
and $\chi_{\text{ag}}(t,t_w)$
is the off equilibrium extra response.
Notice that~(\ref{FDT}) may be read as the statement that 
$\chi_{\text{st}}(t-t_w)$ depends on time through $G_{\text{st}}(t-t_w)$.
Mean field theory for glassy systems predicts\cite{Cugliandolo93} that
in the asymptotic time region this holds also off equilibrium with
$\chi (t,t_w)=\widehat{\chi} (G(t,t_w))$. If this is the case and if
$\lim_{t \rightarrow \infty} \chi(t,t_w)=\chi_{\text{eq}}$
then static and dynamic properties are connected\cite{Franz98} by the relation
\be
\left. -T\frac{d^2\widehat{\chi}(G)}{dG^2} \right )_{G=q} = P_{\text{eq}}(q)
\label{i1.2}
\ee
where  $\chi_{\text{eq}}$ and $P_{\text{eq}}(q)$ are the linear
susceptibility and the overlap probability distribution\cite{Mezard87}
in the equilibrium state. Applying the scheme to phase ordering we 
should find
\be
\left. -T\frac{d^2\widehat{\chi}(G)}{dG^2} \right )_{G=q} = \delta(q-M^2)
\label{i1.2.bis}
\ee
since the overlap function in the equilibrium state is given by 
the delta function\cite{Mezard87}. On the other hand~(\ref{FDT}) may be rewritten as
$T\chi_{\text{st}}(t-t_w)=[1-(G_{\text{st}}(t-t_w)+M^2)]$ and, if $t_w$ is
sufficiently large, parameterizing time through the full autocorrelation
function we have
\be
T\widehat{\chi}_{\text{st}}(G)=\left\{\begin{array}{ll}
	1-G\qquad\qquad           &\mbox{, for $M^2 \leq G \leq 1$} \\  
	1-M^2         & \mbox{, for $G < M^2$}
    \end{array}\right.
\label{i2.2}
\ee
which gives 
$\left. -T\frac{d^2}{dG^2}\widehat{\chi}_{\text{st}}(G) \right )_{G=q} 
= \delta(q-M^2)$. Hence, for~(\ref{i1.2}) to hold one must necessarily have
\be
\lim_{t_w \rightarrow \infty} \chi(t,t_w)= \widehat{\chi}_{\text{st}}(G)
\label{i2.2.1}
\ee
and $\lim_{t_w \rightarrow \infty} \chi_{\text{ag}}(t,t_w)=0$.
Numerical simulations for the Ising model\cite{Barrat98} in $d=2$ and $d=3$
as well as analytical results for the spherical model\cite{Cugliandolo95}
indeed show evidence that~(\ref{i2.2.1}) is asymptotically satisfied.
More precisely, one expects\cite{Berthier99,Parisi99}
the off equilibrium contribution to the response function to scale as
$\chi_{\text{ag}}(t,t_w)= t_w^{-a}\widetilde{\chi}_{\text{ag}}(t/t_w)$
with $a=1/2$ on the basis of the argument that $\chi_{\text{ag}}(t,t_w)$
is proportional to the interface density
$\rho_I(t)\sim L^{-1}(t)$ where $L(t) \sim t^{1/z}$ is the 
typical domain size and $z=2$ for non conserved order parameter  
dynamics\cite{Bray94}.
  
Motivated by analytical results\cite{Lippiello00} for the one dimensional Ising
model which, instead, give $a=0$, recently a detailed 
study of the behavior of the response function has been 
undertaken\cite{Corberi01}
finding an interesting dependence of the exponent $a$ on space dimensionality.
This is best explained in terms of the effective
response $\chi_{\text{eff}}(t,t_w)$ due to a single interface and defined by
\be
\chi_{\text{ag}}(t,t_w) = \rho_I(t)\chi _{\text{eff}}(t,t_w).
\label{i5}
\ee
In this form the behavior of $\chi_{\text{ag}}(t,t_w)$ 
appears to be determined by the
balance between the interface loss due to coarsening and the response 
associated to a single interface. If one requires 
$\chi_{\text{ag}}(t,t_w) \sim \rho_I(t)$
clearly $\chi_{\text{eff}}(t,t_w)$ must be a constant.
In Ref.\cite{Corberi01} it was found that for an Ising system this is the 
case only for
$d>3$, while for $d<3$ there is the power law growth
\be
\chi_{\text{eff}}(t,t_w) \sim (t-t_w)^{\alpha}
\label{i6}
\ee
with numerical values for the exponent at different dimensionalities 
compatible with $\alpha =
(3-d)/4$. At $d=3$ the power law is replaced by logarithmic growth.
It has been conjectured that this dimensionality dependence of 
$\alpha$ is the outcome of the competition between the curvature
of interfaces and the external perturbing field in the drive of
interface motion. According to this picture $d=3$ is the dimensionality 
above which interface
motion is dominated by the curvature, while below $d=3$ the
external field competes effectively with the curvature. The more so
the lower is the dimensionality, until in $d=1$ interfaces reduce to
pointlike objects
driven only by the external field. When this happens, the rate of growth
of the single interface response matches exactly the rate of loss of the
interface density, with $\alpha=1/2$ and $a=0$. Then,
$\chi_{\text{ag}}(t,t_w)$ does not vanish as $t_w \rightarrow \infty$
and the second term in~(\ref{i2}) contributes to $\widehat{\chi}(G)$
producing the violation of~(\ref{i1.2}). 
In short, dimensionality acts as the control parameter which allows to
modulate the competition between the two opposing mechanisms driving interface
motion and summarized by
\be
a=\left\{\begin{array}{ll}
	\frac{1}{2}\qquad\qquad           &\mbox{, for $d > 3$} \\  
	\frac{d-1}{4}         & \mbox{, for $d \leq 3$.}
    \end{array}\right.
\label{i7}
\ee

In this paper we investigate a generalization of this phenomenon occurring 
in the
framework of the random field Ising model (RFIM), which further 
clarifies how the overall behavior of $\chi_{\text{ag}}(t,t_w)$ originates in 
the interplay between the rate of growth of the single 
interface response function and the rate of loss of defect 
density. Here we specialize to the $d=1$ RFIM. In this case  domain 
walls perform random walks
in a random potential of the Sinai type\cite{Fisher01}. The crucial feature
of this walk, from our perspective, is the crossover between
the preasymptotic regime, characterized by free diffusion as in the
pure system, and the asymptotic regime characterized by Sinai diffusion.
Measuring $\chi_{\text{eff}}(t,t_w)$
and $\rho_I(t)$, in the preasymptotic regime we shall find
\be
\chi_{\text{eff}}(t,t_w) \geq \rho_I^{-1}(t)
\label{i8}
\ee
as in the pure system. Namely, as long as diffusion takes place in a flat
landscape the rate of growth of $\chi_{\text{eff}}(t,t_w)$
does not go below the rate
of loss of interfaces. However, going over to the asymptotic regime
the landscape becomes rugged and activated processes do play a role.
When this happens $\chi_{\text{eff}}(t,t_w)$ slows down with
respect to $\rho_I(t)$ eventually reaching
\be
\chi_{\text{eff}}(t,t_w) \sim \rho_I^{-1/2}(t)
\label{i8.1}
\ee
which leads to the vanishing of $\chi_{\text{ag}}(t,t_w)$. Through 
this mechanism the validity of~(\ref{i1.2})  is restored in the $d=1$ RFIM.

\section{Unperturbed system}

In the following we consider the $d=1$
Ising model with Hamiltonian
\be
{\cal H}[\sigma_i] = -J \sum_{i=1}^{N} \sigma_i \sigma_{i+1}
-\sum_{i=1}^{N} h_i \sigma_i
\label{2.1}
\ee
where $J>0$ is the ferromagnetic coupling and $h_i= \pm h_0$
is an uncorrelated random field with expectations
\be
\left \{ \begin{array}{ll}
			E_h(h_i) =0             \\  
			E_h(h_ih_j)= h_0^2\delta_{ij}.    
                   \end{array}                                        
               \right.
\label{2.2}
\ee
Here $h_i$ is not a perturbing field, so $h_0$ does not
have to be small.
We will consider the two different dynamical evolutions taking place
when the system is quenched to the final temperature $T$ starting from the
two initial states:

{\it i}) the system is prepared into a spin configuration containing a
single interface
\be
\sigma_i = \mbox{sgn}(i)
\label{i01}
\ee

{\it ii}) the system is in equilibrium at infinite temperature 
with the uniform measure
\be
{\cal P}_0[\sigma_i]= 2^{-N}.
\label{i02}
\ee
In order to have a phase ordering process the equilibrium state at the
final temperature of the quench must be ordered. 
In the $d=1$ RFIM this is 
not possible even at $T=0$, since the size of the ordered domains is 
limited by the Imry Ma length $L_{IM} = \frac{4J^2}{h_0^2}$. So, we take the
limit\cite{Corberi01} of an infinite ferromagnetic coupling $J \rightarrow \infty$ in
order to have $L_{IM} = \infty$.
In this case the equilibrium state is the mixture of the two
ordered states
\be
{\cal P}_{\text{eq}}[\sigma_i]= \frac{1}{Z}e^{-\frac{1}{T}{\cal H}[\sigma_i]}=
\frac{1}{2}\prod_i \delta(\sigma_i-1)+\frac{1}{2}\prod_i \delta(\sigma_i+1)
\label{2.3}
\ee
for any finite temperature $T$. Notice that in the corresponding pure
states $M^2=1$ which implies $\chi_{\text{eq}}=0$ and 
$P_{\text{eq}}(q)=\delta(q-1)$. 
Furthermore, with $J=\infty$ thermal
fluctuations within domains are suppressed and dynamics is reduced to
interface diffusion. With
the initial condition~(\ref{i01}) we have the diffusion
of the single interface and with the initial condition~(\ref{i02}) the
diffusion-annihilation process of the set of interfaces seeded by the
initial condition. Characteristic lengths of the process are in the first 
case the root mean square displacement (RMSD) $R(t)$ of the single interface
and in the second case the average distance between interfaces (or
average domain size) $L(t)$. Furthermore, 
since the typical size of the potential barrier
that a walker must overcome after traveling a distance $l$ is $l^{1/2}h_0$,
there remains defined another characteristic length
\be
L_{g} = \left (\frac{T}{h_0} \right )^2
\label{2.4}
\ee
as the distance over which potential barriers are of the order of magnitude
of thermal energy. This length is important because separates the distances
smaller than $L_g$, over which diffusion takes place as in the pure system
with $R(t)$ or $L(t) \sim t^{1/2}$, from the distances larger than $L_g$, over
which diffusion is dominated by the random potential and it is of the
Sinai type\cite{Sinai82} with $R(t)$ or $L(t) \sim (\log t)^2$. Clearly, the limit of
the pure system corresponds to $L_g=\infty$.

Considering different temperatures $T$ in the presence of random fields
of different strength $h_0$, we have let the system evolve with initial
conditions~(\ref{i01}) and~(\ref{i02}).
One time unit is defined equal to one flip attempt per spin, on the average
(spins are updated in random order). We use Metropolis transition rates,
that is the probability to flip the spin is
$p_{\text{flip}}=\min(1,\exp(-\frac{1}{T}\Delta{\cal H}))$.
In the case of initial condition~(\ref{i01}),
we have found that
the RMSD of the single interface depends on
$T$ and $h_0$ through $L_g$ satisfying (Fig.~\ref{fig:1})
the scaling relation
\be
R(t,L_g)=L_g {\cal R}\left (\frac{t}{L_g^2} \right )
\label{2.7}
\ee
with
\be
{\cal R}(x) \sim \left \{ \begin{array}{ll}
			x^{\frac{1}{2}}    & \mbox{for $x \ll 1$} \\  
			(\log x)^2\qquad\qquad         & \mbox{for $x \gg 1$}
                   \end{array}                                        
               \right.
\label{2.8}
\ee
which displays the crossover from the pure regime to the Sinai regime. 
A completely analogous behavior is obtained in the second case for the 
typical domain size
$L(t)$, measured as
the inverse density of interfaces $\rho_I^{-1}(t)$.
We find (Fig.~\ref{fig:2}) 
\be
L(t,L_g)=L_g {\cal L}\left (\frac{t}{L_g^2} \right )
\label{2.8.1}
\ee
where the scaling function ${\cal L}(x)$ obeys the same limiting
behaviors as in~(\ref{2.8}).

Next, let us consider the autocorrelation function
$G(t,t_w,L_g)=\frac{1}{N}E_h \left [\sum_i \langle \sigma_i(t)
\sigma_i(t_w)\rangle_h \right ]$
where the angular brackets denote the average over thermal noise 
for a given realization of the field $[h_i]$.
Due to the absence of thermal fluctuations within domains
$G_{\text{st}}(t-t_w)  \equiv 0$ and the autocorrelation function is
entirely given by the aging component which satisfies (Fig.~\ref{fig:3}) the
scaling relation
\be
G_{\text{ag}}(t,t_w,L_g)= F\left (\frac{L(t)}{L(t_w)},z \right )
\label{2.9}
\ee
where $z=\frac{L(t_w)}{L_g}$. The shape of the scaling function $F(x,z)$
is known in the limits of the pure system $(z=0)$ and of the Sinai regime
$(z=\infty)$. In the first case from~(\ref{2.8.1}) follows 
$x=\sqrt{\frac{t}{t_w}}$ and from the exact solution\cite{Bray89} of the
Glauber dynamics for the Ising chain
\be
F(x,z=0)= \frac{2}{\pi} \arcsin \left (\frac{2}{1+x^2} \right ).
\label{2.11}
\ee
The second case is obtained taking $t_w$ so large that $\ln L_g$ can
be neglected with respect to $\ln t_w$. This yields 
$x=\left ( \frac{\ln t}{\ln t_w}\right )^2$ and\cite{Fisher01}
\be
F(x,z=\infty) = \frac{4}{3 \sqrt{x}}- \frac{1}{3x}.
\label{2.12}
\ee
In the intermediate cases with finite values of $z$ one expects
\be
F(x,z)  =     \left \{ \begin{array}{ll}
			F(x,0)    & \mbox{, for $x-1 \ll 1/z$} \\  
			F(x,\infty)       & \mbox{, for $x-1 \gg 1/z$}
                   \end{array}                                        
               \right.
\label{2.23}
\ee
where the condition on $x$ can be rewritten more transparently 
as $L(t)-L(t_w) \ll L_g$ or $L(t)-L(t_w) \gg L_g$. While our data in
Fig.~\ref{fig:3}
show clearly that~(\ref{2.23}) is obeyed for $x-1 \ll 1/z$, the check 
of the crossover to the Sinai regime requires simulation times
too long for what we can achieve. In any case, the behavior of the data 
under variation of $z$ in Fig.~\ref{fig:3} shows clearly the onset
of the crossover. Notice that~(\ref{2.23}) means that even if the 
shortest time $t_w$ is chosen inside the Sinai regime, namely
after the potential has developed a rough
landscape, for displacements up to $L_g$ in the bottom of the potential 
valleys interface diffusion takes place as in the pure system.

Finally, let us consider the behavior of the staggered magnetization
\be
M(t,L_g)=\frac{T}{Nh_0^2}
E_h \left [ \sum_i \langle \sigma_i(t)\rangle_h h_i \right ].
\label{2.10.1}
\ee
Essentially, this quantity gives the behavior of (minus) the magnetic
energy per spin. At $t=0$ the spin and field configurations are uncorrelated
and $M(t=0,L_g)=0$. As time evolves one expects the spins to correlate with the
field producing growth in $M(t,L_g)$. 
Indeed, $M(t,L_g)$ displays (Fig.~\ref{fig:4}) growth in the 
preasymptotic regime $L(t) \ll L_g$, which however is followed by 
an intermediate regime with a constant plateau for $L(t) \sim L_g$ and then 
by a decrease toward zero for $L(t) \gg L_g$. In the intermediate and
asymptotic regimes one has
\be
M(t,L_g)= {\cal M}\left (\frac{L(t)}{L_g} \right )
\label{2.10.2}
\ee
with ${\cal M}(x) \sim x^{-1/2}$ for $x \gg 1$.
In order to understand how this comes about it is useful to look at
$M(t,L_g)$ as made up by separate contributions associated to the single
interfaces whose number decreases as time goes on. This is formalized by
writing
\be
M(t,L_g)=\rho_I(t) M_{\text{eff}}(t,L_g)
\label{2.10.3}
\ee
which defines $M_{\text{eff}}(t,L_g)$ as the effective staggered magnetization
associated to a single interface. Using~(\ref{2.10.2}) then we have
\be
M_{\text{eff}}(t,L_g) = L_g{\cal M}_{\text{eff}}
\left (\frac{L(t)}{L_g} \right )
\label{2.10.4}
\ee
with (Fig.~\ref{fig:5}) the function ${\cal M}_{\text{eff}}(x)$ obeying
\be
{\cal M}_{\text{eff}}(x) \sim \left\{\begin{array}{ll}
	x\qquad\qquad       & \mbox{for $x \ll 1$} \\  
		\sqrt{x}        & \mbox{for $x \gg 1$.}
                   \end{array}                                        
               \right.
\label{2.15}
\ee
The interpretation of $M_{\text{eff}}(t,L_g)$ as the single interface
contribution to the build up of magnetization can be substantiated by
measuring $M(t,L_g)$ in the process with the initial condition~(\ref{i01}).
In this case, defining $M_{\text{single}}(t,L_g)=NM(t,L_g)$, we find
(Fig.~\ref{fig:6})
\be
M_{\text{single}}(t,L_g)= L_g{\cal M}_{\text{single}}
\left (\frac{L(t)}{L_g} \right )
\label{2.10.5}
\ee
where ${\cal M}_{\text{single}}(x)$ displays the behavior~(\ref{2.15}). From
this we may draw the following conclusions: {\it i}) when in the system 
there is only one interface, the spin-field correlation grows with time.
Therefore, this is not a local effect involving only the alignment of
the pair of spins at the interface with the local field. Rather, it is a
large scale effect involving the optimization of the interface position
with respect to the entire field configuration. {\it ii}) The magnetization
growth takes place with different time laws in the preasymptotic and
asymptotic regimes. {\it iii}) The total magnetization behavior of
Fig.~\ref{fig:4},
when multiple interfaces are present, then is just due to the fact that in the
preasymptotic regime the rate of growth of the single interface magnetization
balances the rate of loss of interfaces, while in the asymptotic regime
interfaces do disappear faster than the growth of the single interface
contribution to the magnetization.

\section{Response function}

Let us  now consider what happens if at the time $t_w >0$ an additional
random field $\epsilon_i = \pm \epsilon_0$ uncorrelated with $h_i$ and 
with expectations
\be
\left \{ \begin{array}{ll}
	          E_{\epsilon}(\epsilon_i)=0       \\  
                  E_{\epsilon}(\epsilon_i \epsilon_j)=\epsilon_0^2\delta_{ij} 
                   \end{array}                                        
               \right .
\label{3.1}
\ee
is switched on.  We take
$\epsilon_0 \ll h_0$ and we are interested in the linear response function
with respect to the $\epsilon$-perturbation given by
\be
T \chi(t,t_w,L_g) = \frac{T}{N\epsilon_0^2}E_hE_{\epsilon} \left [
\sum_i \langle \sigma_i(t) \rangle_{h,\epsilon} \epsilon_i \right ]
\label{3.6}
\ee
where the external field $h$ acts from $t=0$ and undergoes the change 
from $h$ to $h+\epsilon$ at $t_w$.

Due to the suppression of thermal fluctuations within
domains enforced through $J=\infty$, the stationary contribution
$\chi_{\text{st}}(t-t_w)$ 
in~(\ref{i2}) vanishes identically. Therefore, the above response function
is entirely constituted by the aging term $\chi_{\text{ag}}(t,t_w)$.
The interest, then, is focused on the scaling properties. As in the case
of the autocorrelation function, we find (Fig.~\ref{fig:7}) that this quantity
obeys the scaling form
\be
T \chi(t,t_w,L_g)= \widetilde{\chi}
\left ( \frac{L(t)}{L(t_w)},z \right ).
\label{3.2.1}
\ee
For $z=0$ the pure system response function\cite{Lippiello00} is recovered
\be
\widetilde{\chi}(x,z=0) = \frac{\sqrt{2}}{\pi} \arctan \sqrt{x^2-1}
\label{3.8}
\ee
which is responsible of the violation of~(\ref{i1.2})
in the pure Ising chain\cite{Corberi01}. In fact, 
from the equilibrium state~(\ref{2.3})
we ought to have $P_{\text{eq}}(q)=\delta(q-1)$ which is consistent 
with~(\ref{i1.2}), as explained in the  Introduction,
if only $\widehat{\chi}_{\text{st}}(G)$ given by~(\ref{i2.2}) enters into
$\widehat{\chi}(G)$ and the limit $M^2 \rightarrow 1$ is taken 
after differentiation. Instead, eliminating $x$ between~(\ref{3.8}) 
and~(\ref{2.11}) one finds 
\be
\widehat{\chi}(G,z=0) = 
      \frac{\sqrt{2}}{\pi}\arctan [\sqrt{2} \cot (\frac{\pi}{2}G)]
\label{3.9}
\ee
which spoils~(\ref{i1.2}).

With $z>0$ there is a crossover. For values of $x$ up to $x-1 \sim 1/z$
Fig.~\ref{fig:7}
shows that $\widetilde{\chi}(x,z)$ behaves as in the pure case, on the
basis of the same argument used for the autocorrelation
function. For larger values of $x$ the data show that $\widetilde{\chi}(x,z)$
levels off and then decreases. This is clearly displayed also in the 
plot (Fig.~\ref{fig:8})
against the autocorrelation function. The mechanism responsible
of this behavior is the same discussed in the previous Section for the
staggered magnetization. Let us look at the effective response of a
single interface $\chi_{\text{eff}}(t,t_w)$ defined in~(\ref{i5}). 
We find
\be
T\chi_{\text{eff}}(t,t_w,L_g) = L(t_w)
\widetilde{\chi}_{\text{eff}}
\left ( \frac{L(t)}{L(t_w)},z \right )
\label{3.9.1}
\ee
with the scaling function displaying (Fig.~\ref{fig:9}) the behavior
\be
\widetilde{\chi}_{\text{eff}}(x,z) \sim \left\{\begin{array}{ll}
	\widetilde{\chi}_{\text{eff}}(x,z=0)\qquad\qquad       
              & \mbox{, for $x-1 \ll 1/z$} \\  
		\sqrt{x}        & \mbox{, for $x-1 \gg 1/z$}
                   \end{array}                                        
               \right.
\label{3.9.2}
\ee
where $\widetilde{\chi}_{\text{eff}}(x,z=0)=x\widetilde{\chi}(x,z=0)$.
>From this follows~(\ref{i8}) in the preasymptotic regime and~(\ref{i8.1})
in the asymptotic regime, which account for the crossover of the
response function in Fig.~\ref{fig:7} in terms of the balance between the rate
of growth of the single interface response and the rate of loss of
interfaces. Hence, for $z>0$ eventually $\widetilde{\chi}(x,z)$ 
vanishes and in the limit $z \rightarrow \infty$ one
expects
\be
\widetilde{\chi}(x,z=\infty) \equiv 0.
\label{3.9.3}
\ee
Therefore, for any finite $h_0$ the validity of~(\ref{i1.2}) is restored
asymptotically.

Lastly, it remains to make sure that $\widetilde{\chi}_{\text{eff}}(x,z)$
may be identified with the response function of a single interface.
This we have done by measuring the response function
$\chi_{\text{single}}(t,t_w,L_g)$ in the process with only one
interface in the initial condition and finding (Fig.~\ref{fig:10}) a behavior
quite close to the one of Fig.~\ref{fig:9}.

\section{Conclusions}

One of the hypothesis for the validity of~(\ref{i1.2}) relating static
and dynamic properties is that the large time limit of
$\chi(t,t_w)$ reaches the equilibrium value $\chi_{\text{eq}}$. 
It is then 
clear from~(\ref{i2}) that for this to be true in the phase ordering
process $\chi_{\text{ag}}(t,t_w)$ must vanish, since this is an
intrinsically out of equilibrium contribution.
In other words, the existence of the interfacial degrees of freedom
which do not equilibrate must not play a role at the level of the
response function. Indeed,
$\chi_{\text{ag}}(t,t_w)$ normally does to vanish. However, there
are exceptions in special cases and the study of these is quite instructive
since it allows to gain insight into the properties of
$\chi_{\text{ag}}(t,t_w)$. One dimensional systems with a scalar order
parameter and frozen bulk thermal fluctuations do make such a special
case. In the $d=1$ pure Ising model, due to the pointlike structure of
domain walls, the minimization of magnetic energy introduces a bias
in the diffusion of interfaces which leads to a nonvanishing
$\chi_{\text{ag}}(t,t_w)$ through~(\ref{i8}).
Elsewhere\cite{Corberi01} we have analyzed how~(\ref{i8}) ceases to hold
in going from $d=1$ to $d>1$, due to the extended nature of interfaces.
In that case the minimization of magnetic energy is hindered by the 
competing need to minimize the curvature
of interfaces and $\chi_{\text{ag}}(t,t_w)$ asymptotically disappears.
In this paper we have investigated a different mechanism altering 
the delicate balance~(\ref{i8}) between the gain and loss of contributions
to the response function, which does not require the passage from
pointlike to extended defects. In the $d=1$ RFIM it is the gradual roughening
of the landscape that slows down the minimization of the magnetic energy
with respect to the growth law of the domain size and which eventually
yields an asymptotically vanishing $\chi_{\text{ag}}(t,t_w)$.
Therefore, the overall picture which comes out is that a coarsening
system is always out of equilibrium in the sense that there are
always interfaces around, each one producing the out of equilibrium 
contribution $\chi_{\text{eff}}(t,t_w)$ to the response function,
independently from dimensionality or the presence of quenched disorder.
These are elements which affect the finer details, like whether
$\chi_{\text{eff}}(t,t_w)$ is a constant or how fast it grows.
Where these properties come into play is in putting together the
contributions of all the separate interfaces through
$\chi_{\text{ag}}(t,t_w) = \rho_I(t)\chi_{\text{eff}}(t,t_w)$ which, now
depending on dimensionality or quenched disorder, may then produce
either a vanishing or an asymptotically persistent net result.

Finally, a comment should be made about the non commutativity of the
order of the limits $h_0 \rightarrow 0$ and $t_w \rightarrow \infty$.
If the limit $h_0 \rightarrow 0$ is taken first, the linear response 
function of the pure Ising model is obtained and, as discussed above,
(\ref{i1.2}) is violated. Instead, if $t_w \rightarrow \infty$ is
taken first one finds the asymptotic linear response function~(\ref{3.9.3})
of the RFIM which is consistent with~(\ref{i1.2}). However, if the limit
$h_0 \rightarrow 0$ is taken next, the response function sticks to
$\widetilde{\chi}(x,z=\infty)$, which is not the linear response function
of the pure model.

Acknowledgments - This work has been partially supported
from the European TMR Network-Fractals c.n. FMRXCT980183 and 
from MURST through PRIN-2000.

\def\figsize{8cm}

\begin{figure}
\begin{center}
\mbox{\epsfysize=\figsize\epsfbox{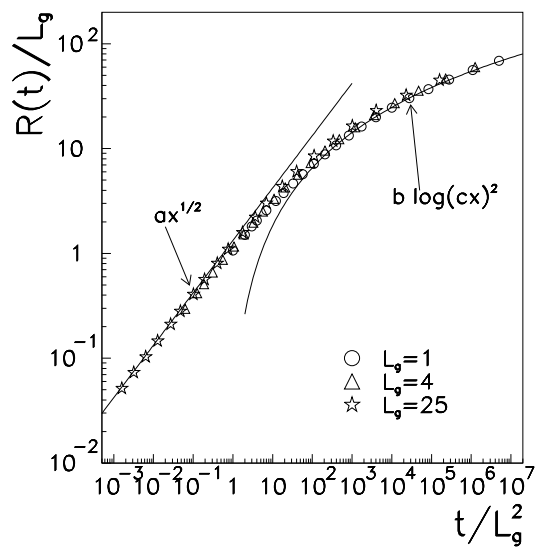}}
\end{center}
\caption{Rescaled
RMSD $R(t)/L_g$
versus rescaled time $t/L_g^2$, for a single interface and $L_g=1,4,25$.
The fit parameters are $a=1.33$, $b=0.274$ and $c=1.33$.
}
\label{fig:1}
\end{figure}

\begin{figure}
\begin{center}
\mbox{\epsfysize=\figsize\epsfbox{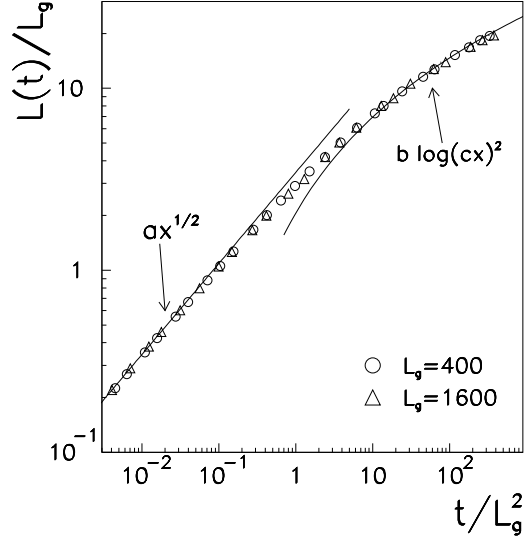}}
\end{center}
\caption{Rescaled average length of domains $L(t)/L_g$ versus
rescaled time $t/L_g^2$, for multiple interfaces and
$L_g=400,1600$.
The fit parameters are $a=3.45$, $b=0.271$ and $c=15.8$.
}
\label{fig:2}
\end{figure}

\begin{figure}
\begin{center}
\mbox{\epsfysize=\figsize\epsfbox{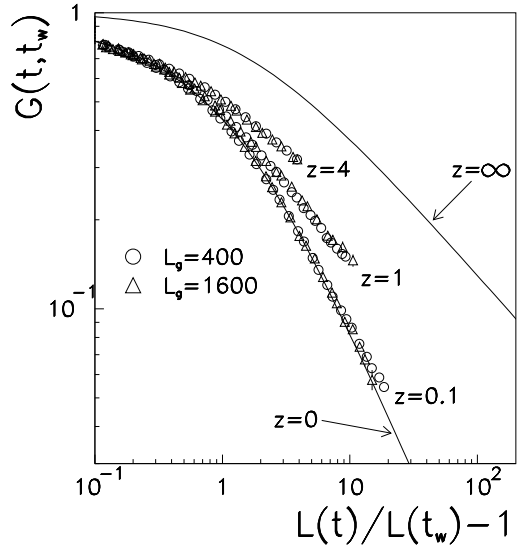}}
\end{center}
\caption{Autocorrelation function $G(t,t_w)$
versus the ratio $(L(t)-L(t_w))/L(t_w)$, 
for $L_g=400$, 1600, and $z=0.1$, 1, 4, with $z=L(t_w)/L_g$. 
Solid lines are the exact results
for $z\to 0$ and $z\to\infty$.}
\label{fig:3}
\end{figure}

\begin{figure}
\begin{center}
\mbox{\epsfysize=\figsize\epsfbox{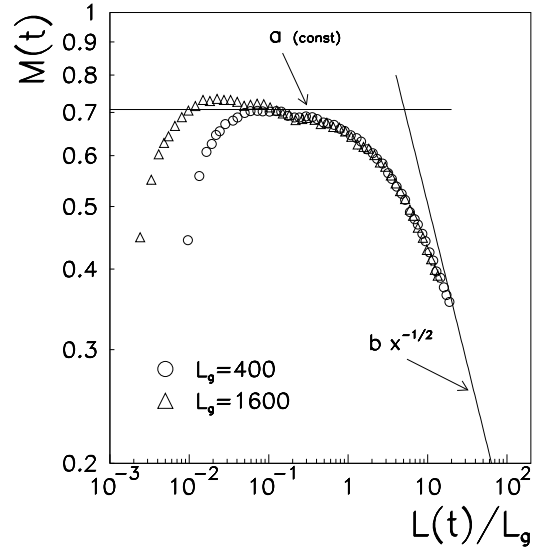}}
\end{center}
\caption{Total staggered magnetization $M(t)$ versus
rescaled average length
of domains $L(t)/L_g$, for $L_g=400$, 1600.}
\label{fig:4}
\end{figure}

\begin{figure}
\begin{center}
\mbox{\epsfysize=\figsize\epsfbox{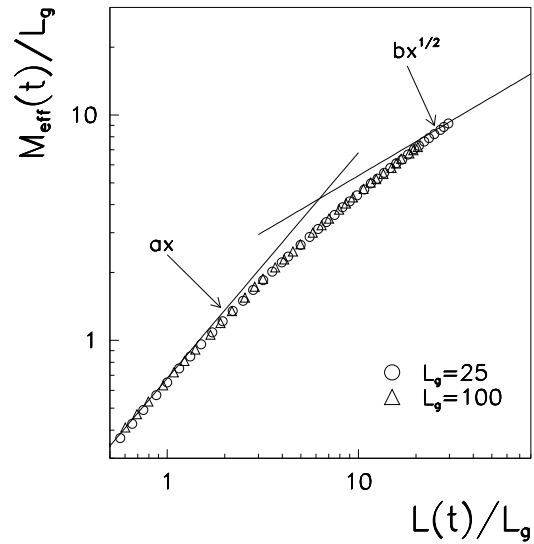}}
\end{center}
\caption{Rescaled effective staggered magnetization $M_{\text{eff}}(t)/L_g$
versus rescaled average length of domains $L(t)/L_g$, for $L_g=25$, 100.
The fit parameters are  $a=0.68$ and $b=1.7$.
}
\label{fig:5}
\end{figure}

\begin{figure}
\begin{center}
\mbox{\epsfysize=\figsize\epsfbox{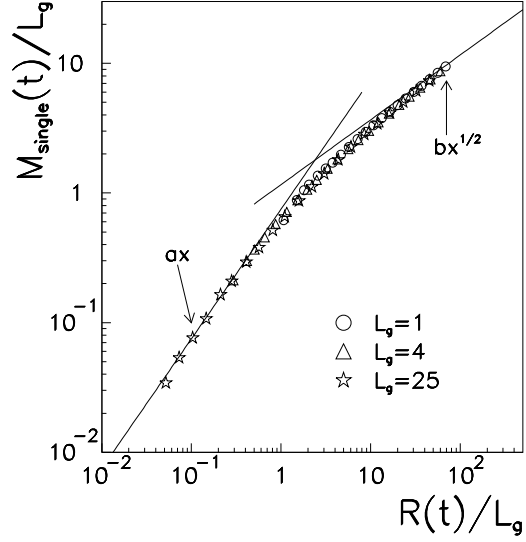}}
\end{center}
\caption{Rescaled single interface staggered magnetization
$M_{\text{single}}(t)/L_g$
versus rescaled RMSD $R(t)/L_g$, for $L_g=1,4,25$.
The fit parameters are $a=0.75$ and $b=1.15$.
}
\label{fig:6}
\end{figure}

\begin{figure}
\begin{center}
\mbox{\epsfysize=\figsize\epsfbox{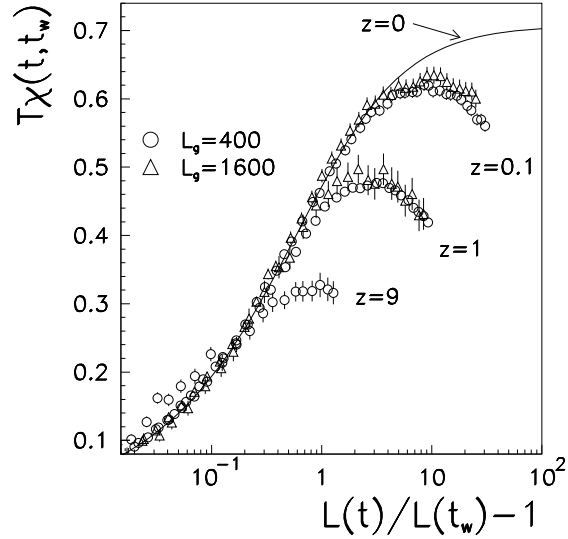}}
\end{center}
\caption{Total linear response $T\chi(t,t_w)$ versus the ratio
$(L(t)-L(t_w))/L(t_w)$, for $L_g=400$, 1600, and $z=0.1$, 1, 9. 
The solid line is
the exact result for $z\to 0$.}
\label{fig:7}
\end{figure}

\begin{figure}
\begin{center}
\mbox{\epsfysize=\figsize\epsfbox{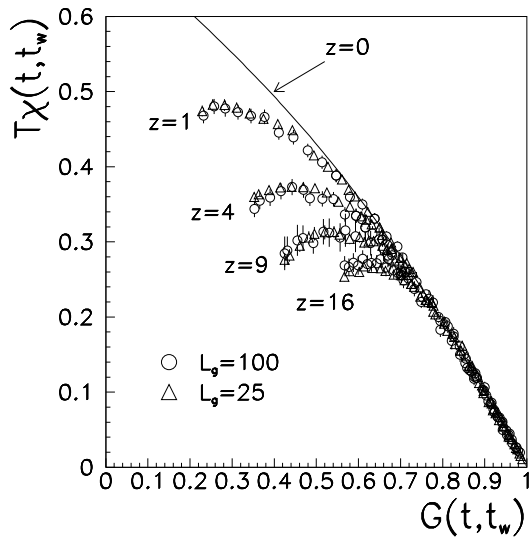}}
\end{center}
\caption{Total linear response $T\chi(t,t_w)$ versus
autocorrelation $G(t,t_w)$,
for $L_g=25,100$ and $z=1,4,9,16$. The solid line is the exact
result for $z\to 0$.}
\label{fig:8}
\end{figure}

\begin{figure}
\begin{center}
\mbox{\epsfysize=\figsize\epsfbox{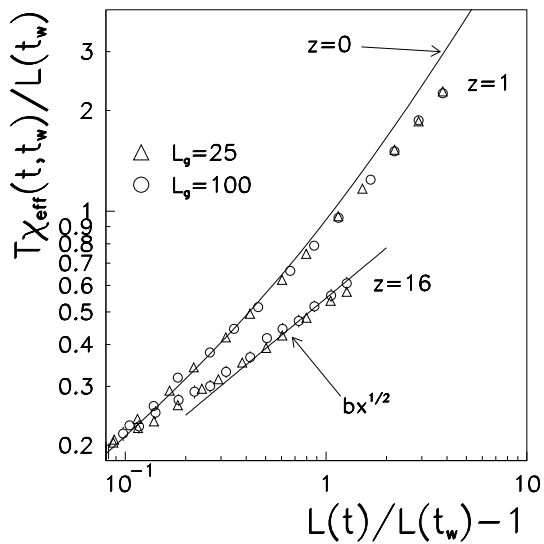}}
\end{center}
\caption{Rescaled effective linear response
$T\chi_{\text{eff}}(t,t_w)/L(t_w)$ versus the ratio
$(L(t)-L(t_w))/L(t_w)$, for $L_g=25$, 100, and $z=1$, 16.
The solid line is the exact result for $z\to 0$.}
\label{fig:9}
\end{figure}

\begin{figure}
\begin{center}
\mbox{\epsfysize=\figsize\epsfbox{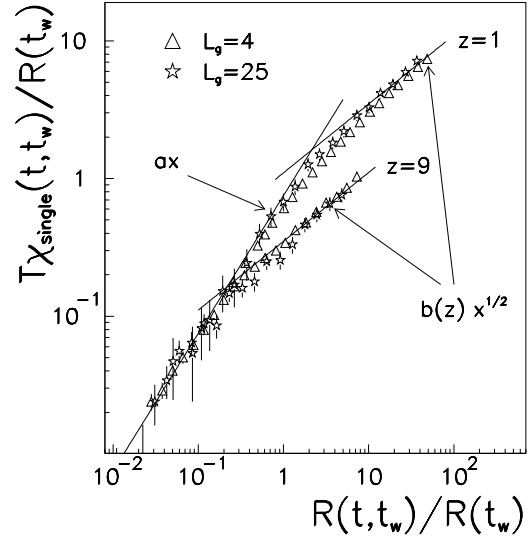}}
\end{center}
\caption{Rescaled single interface linear response
$T\chi_{\text{single}}(t,t_w)/R(t_w)$
versus the ratio $R(t,t_w)/R(t_w)$, where $R(t,t_w)$ and $R(t_w)$ 
are respectively
the RMSDs between times $t_w$ and $t$, and between times 0 and $t_w$,
for $L_g=4$, 25, and $z=1$, 9.
The fit parameters are 
$a=0.75$, $b(1)=1.1$, $b(9)=0.35$.
}
\label{fig:10}
\end{figure}

%\end{multicols}
\end{document}